\documentclass[prb,aps,showpacs,floatfix,twocolumn]{revtex4}
\usepackage{epsfig}
\usepackage{amsmath}
\usepackage{amsfonts}
\usepackage{amssymb}
\usepackage{mathptmx}
\usepackage{eucal}

\DeclareMathOperator{\im}{\mathop{\mathrm{Im}}}
\DeclareMathOperator{\const}{\mathop{\mathrm{const}}}
\newcommand{\Eq}[1]{Eq.~(\ref{#1})}
\newcommand{\Eqs}[1]{Eqs.~(\ref{#1})}

\begin{document}
\title{Electric potential of the electron sound wave: Sharp
disappearance in the superconducting state}

\author{Yu.~A.~Avramenko, E.~V.~Bezuglyi, N.~G.~Burma, and V.~D.~Fil}

\affiliation{B.Verkin Institute for Low Temperature Physics and Engineering,
National Academy of Sciences of Ukraine, 47 Lenin Ave., Kharkov UA-61103,
Ukraine}

\begin{abstract}
We study the ac electric potential induced by the electron sound wave (a
perturbation of the electron distribution function propagating with the Fermi
velocity) in single crystals of high purity gallium. The potential and the
elastic components of the electron sound demonstrate qualitatively different
dependencies on the electron relaxation rate: while the phase of the potential
increases with temperature, the phase of elastic displacement decreases. This
effect is explained within the multiband model, in which the potential is
attributed to the ballistic quasiwave, while the elastic component is
associated with the zero-sound wave. We observed a mysterious property of the
superconducting state: all manifestations of the potential accompanying the
lattice deformations, including usual sound wave, disappear below
$T_{\text{c}}$ in almost jumplike manner.
\end{abstract}


\maketitle

\section{Introduction}\vspace{-0.3cm}

Longitudinal perturbations of the electronic and elastic subsystems of metals
are accompanied by fluctuations of the electron density and occurrence of
alternating electric fields, which provide electrical neutrality of the system.
The ac electric potential of a longitudinal sound wave has been first measured
in Ref.~\onlinecite{AvrGokh}; it should be distinguished from the nonlinear dc
potential arising due to the dragging effect.\cite{Parm,Wein,Zav} In metals,
besides the acoustic mode, there exists several types of electron sound, i.e.,
longitudinal oscillations of the electron distribution function, propagating
with nearly Fermi velocity and coupled to elastic deformations and electric
fields: acoustic plasmons,\cite{Pines} ballistic quasiwaves,\cite{Ivanovski}
zero sound.\cite{Gorkov,Dunin,Dubovik,ZS,BezPhysicaB} The study of these fast
modes gives important information about the mechanisms of electron relaxation,
spectrum of the Fermi velocities, and Fermi-liquid correlation function in
metals.\cite{ZS,BezPhysicaB,BezCondMatter,BezJLTP,Avr,BezVelocity}

In the experiments mentioned above, excitation of the electron sound has been
performed by a high-frequency elastic deformation of the sample surface. As a
result, both the acoustic and the electron sound waves were simultaneously
excited in the bulk of the sample. These waves can be easily separated in the
time-of-flight experiment; the signal $\varphi_\text{S}$, which comes with the
sound delay, will be referred to below as ``sound potential'', and the signal
$\varphi_{\text{ES}}$ propagating with the Fermi velocity will be called
``electron sound potential''. It should be noted, however, that the potential
as well as the elastic displacement are measured at the metal boundary, where
partial conversion between different types of the oscillations always occurs.
Therefore the recorded signal is generally the result of interference between
different processes, and its magnitude may differ from its bulk value in the
propagating wave. Indeed, an analysis of the electric signals of the first
type, assuming specular reflection of electrons from the sample surface,
showed\cite{AvrGokh} that the potential $\varphi_{\text{S}}$ is formed by two
contributions: $\varphi_{\text{q}}$, which has the sound spatial period, and
$\varphi_{\text{qw}}$, generated by the fast quasiwave excited at the receiving
interface. Diffuseness of the sample boundary numerically modifies the effect,
but the main features remain qualitatively unchanged.\cite{Gokh}

Elastic deformations coupled to the electron sound have been studied in
Refs.~\onlinecite{ZS,BezPhysicaB,BezCondMatter,BezJLTP,Avr,BezVelocity} and
\onlinecite{Kopel}. In the present paper, we pay our main attention to the
electron sound potential $\varphi_{\text{ES}}$. This study is of interest due
to the following reasons. First, it provides additional arguments in favor of
earlier assumptions\cite{ZS,BezPhysicaB} of Fermi-liquid nature of the electron
sound, enabling us to separate the zero-sound mode from the quasiwave.
Furthermore, the behavior of $\varphi_{\text{ES}}$ in the superconducting state
is a topic of particular interest. The earlier study of $\varphi_{\text{S}}$
has revealed a quite unexpected effect:\cite{AvrGokh} the sound potential
disappeared almost abruptly below $T_{\text{c}}$. In the present study, we
found a similar effect for the potential $\varphi_{\text{ES}}$ of the electron
sound, it also abruptly disappears below $T_{\text{c}}$, though its elastic
component changes more smoothly in the superconducting state. Such a behavior
of $\varphi_{\text{S}}$ and $\varphi_{\text{ES}}$ has no explanation within the
existing knowledge about the penetration of the longitudinal electric field in
superconductors.\cite{Artemenko} It can be thought that the mysterious behavior
of the electric field generated by an inhomogeneous elastic deformation in a
superconductor is a common property of the superconducting phase irrespective
of investigated materials.

The paper is organized as follows. In Sec.~II, we present the measured
temperature dependencies of the modulus and the phase of the signals. In
Sec.~III, we examine various theoretical models that describe formation of
$\varphi_{\text{ES}}$ for both diffusive and specular interfaces. We conclude
that the theory of elasticity of metals\cite{Kontorovich} applied to multiband
metals satisfactorily describes the behavior of $\varphi_{\text{ES}}$ and
$\varphi_{\text{S}}$ in the normal state. The results for the superconducting
state are presented in Sec.~IV.\vspace{-0.5cm}

\section{Experimental setup and results in the normal state}\vspace{-0.3cm}

The experimental setup was the same as described in Refs.~\onlinecite{AvrGokh}
and \onlinecite{Avr}. One of the faces of a high-pu\-ri\-ty gallium single
crystal (impurity mean free path $\sim $ 5 mm) was excited through the delay
line by a longitudinal elastic wave with the frequency of 55 MHz and the
diameter of the sound beam $\sim $ 4 mm. The elastic component of the signal at
the opposite face of the sample was registered by a piezoelectric transducer,
and the electric potential was measured by an electrode attached to the sample
within the region of the ``sound spot''. The electrode has been made from the
sound-ab\-sor\-bing material (brass) to prevent appearance of its own
potential. In contrast to Ref.~\onlinecite{AvrGokh}, where we used a point
contact, the electrode diameter was $2.5$ mm, which provides more controlled
mechanical boundary conditions. The electrode can be glued to the sample
surface by the acoustic grease or slightly appressed to the sample by a spring.
Obviously, the first case is more similar to a matched boundary, while the
second case to a free one. The experiment was performed in the time-of-flight
regime: the duration of the excitation signal has been chosen smaller than the
sound delay in the sample, which excludes the possibility of the acoustic
resonance. The electron sound resonance was suppressed due to diffusive
scattering of electrons at the sample boundaries.

\begin{figure}[t]
\centerline{\epsfxsize=8.5cm\epsffile{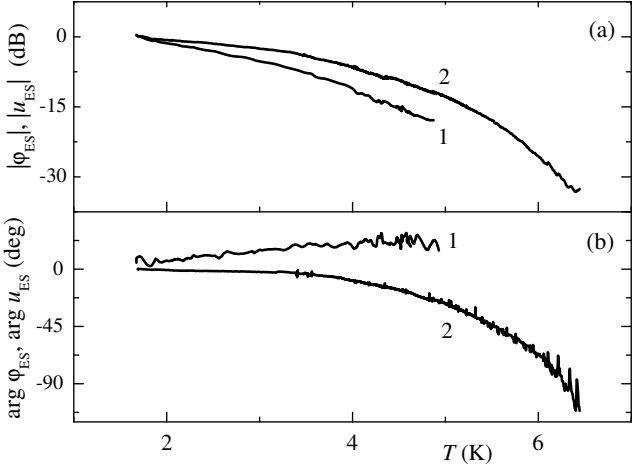}}\vspace{-0.3cm}
\caption{Amplitudes (a) and phases (b) of the potential $\varphi_{\text{ES}}$
(curves 1) and elastic displacement $u_{\text{ES}}$ (curves 2) in the electron
sound wave vs temperature.}\vspace{-0.2cm}
\label{fig1}
\end{figure}
\begin{figure}[t]
\centerline{\epsfxsize=8.5cm\epsffile{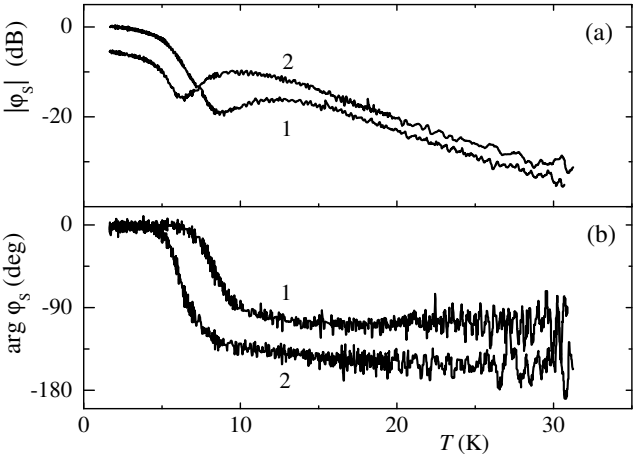}}\vspace{-0.3cm}
\caption{Amplitude (a) and phase (b) of the potential $\varphi_{\text{S}}$ of
the acoustic wave vs temperature; curves 1 is for ``almost free'' interface,
curves 2 is for ``almost matched'' interface.}\vspace{-0.5cm}
\label{fig2}
\end{figure}

In all cases, we detected two types of signals: fast modes propagating with the
Fermi velocity (electron sound) and slow ones having the sound velocity. The
elastic component $u_{\text{ES}}$ of the electron sound was about of 80 dB
lower than its value $u_{\text{S}}$ in the acoustic signal, while the
magnitudes of the potentials $\varphi_{\text{ES}}$ and $\varphi_{\text{S}}$
were comparable. For the sample length of 4 mm in the temperature range of
impurity scattering and excitation intensity $\sim 10$ W/cm$^{2}$, the measured
potentials have the level of $10^{-5}$ V. The amplitude of
$\varphi_{\text{ES}}$ was found to be independent of the mechanical boundary
conditions, while $\varphi_{\text{S}}$ exceeds $\varphi_{\text{ES}}$ by 8 dB
for a ``matched'' boundary and by 14 dB - for an almost free boundary. The
dependencies of the amplitude and phase of $\varphi_{\text{ES}}$ and
$u_{\text{ES}}$ on the temperature (i.e., on the electron scattering rate),
measured in the same sample in the normal state, are shown in Fig.~\ref{fig1}.
We draw one's attention to the qualitative difference between the behavior of
the phases for these components, which seems to be a decisive test for possible
theories, as we will see below. The temperature changes of the amplitude and
phase of $\varphi_{\text{S}}$ are shown in Fig.~\ref{fig2}. In contrast to
analogous data presented in Ref.~\onlinecite{AvrGokh}, the potential
$\varphi_{\text{S}}(T)$ shows a more complicated nonmonotonic behavior, which
indicates interference of nearly antiphase signals, $\varphi_{\text{q}}$ and
$\varphi_{\text{qw}}$, whose amplitudes have different dependencies on the
electron scattering intensity.\cite{AvrGokh}\vspace{-0.5cm}

\section{Theoretical analysis}\vspace{-0.3cm}

\subsection{Free-electron model}\vspace{-0.3cm}

As in Ref.~\onlinecite{AvrGokh}, we first analyze a free-electron model by
using a slightly different approach, which is consistent with the scheme of the
experiment and enables simultaneous calculations of the potentials
$\varphi_{\text{ES}}$ and $\varphi_{\text{S}}$. We consider a metal plate with
the thickness $x_0$, subjected by elastic vibrations with the amplitude $u_0$
at the face $x = 0$, and calculate $\varphi_{\text{ES}}$ and
$\varphi_{\text{S}}$ at $x=x_0$. For simplicity, we assume the same densities
$\rho$ and sound velocities $s$ for the delay line, the receiving electrode and
the sample. The system of equations\cite{Kontorovich} consists of the
one-dimensional linearized kinetic equation in the relaxation time
approximation [the time dependence is chosen as $\exp(i\omega t)$],
\begin{equation} \label{eq1}
i\omega \psi + v\frac{d\psi}{dx} + \nu \psi = - i\omega \Lambda
\frac{du}{dx} + ev\frac{{d\varphi} }{{dx}},
\end{equation}
the equation of the elasticity theory,
\begin{equation} \label{eq2}
-\rho \omega ^2 u = \rho s^2 \frac{d^2 u}{dx^2} - \frac{dW}{dx},\quad W =
\left\langle \Lambda \psi  \right\rangle,
\end{equation}
and the electroneutrality condition
\begin{equation} \label{eq3}
\left\langle \psi \right\rangle = 0.
\end{equation}
Here, $\psi$ is a nonequilibrium addition to the distribution function, $u$ is
the elastic displacement, $v$ is the $x$-component of the Fermi velocity,
$\Lambda$ is the longitudinal part of the deformation potential ($\Lambda =
\lambda - \left\langle {\lambda}  \right\rangle /\left\langle 1 \right\rangle
$, $\lambda = - mv^{2}$), $\nu$ is the relaxation frequency,
\begin{equation} \label{eq4}
\varphi = \varphi_E + \frac{1}{e} \frac{du}{dx}\frac{\left\langle\lambda
\right \rangle}{\left\langle 1 \right\rangle} - \frac{m\omega ^2}{e}\int_0^x
udx
\end{equation}
is a full electrochemical potential measured by a voltmeter, $\varphi_E$ is
its electrical component satisfying Maxwell's equations. The last term in
\Eq{eq4} describes a small Stewart-Tol\-men's effect, which can be ignored
for all cases analyzed below. The angle brackets denote averaging over the
Fermi surface,
\begin{equation} \nonumber
\left\langle A \right\rangle \equiv \frac{2}{h^3}\int \frac{AdS}{v_F}.
\end{equation}
Equations \eqref{eq1} and \eqref{eq3} lead to the condition of the absence of
the longitudinal current,
\begin{equation} \label{eq5}
\left\langle {v\psi}  \right\rangle = 0.
\end{equation}
The condition of completely diffusive reflection of electrons for the
boun\-dary $x = 0$ has the form
\begin{equation} \label{eq6}
\psi (x = 0) = \begin{cases} \psi_0(v), & v < 0 \\ C_0=\const, & v>0,
\end{cases}
\end{equation}
and similarly for $x = x_0$. The function $\psi_0$ and the constant $C_0$
must be found self-consistently.

Typically, such a problem is solved by the representation of the solution of
\Eq{eq1} through an integral formula,\cite{Gokh,Abrikosov} with subsequent
solution of integro-differential equations by the Wie\-ner-Hopf
method.\cite{Noble} We will use a more simple implementation of this method:
extending the solution to the whole $x$-axis and assuming all fields outside
the interval (0, $x_0$) to be zero, we apply the integral Fourier
transformation
\begin{align} \nonumber
A_k = \int_0^{x_0} A(x) \exp(- ikx) dx, \quad A(x) = \int_{-\infty}^{\infty}
A_k \exp(ikx) \frac{dk}{2\pi}
\end{align}
directly to \Eqs{eq1}--\eqref{eq3}. The Fourier transform of the kinetic
equation \eqref{eq1} reads as
\begin{align} \label{eq7}
&\psi_k (L + ik) - \omega k\frac{\Lambda}{v}u_k - iek\varphi_k
\\ \nonumber
&- \left[- i\omega \frac{\Lambda}{v}u (x_0) + e\varphi(x_0) -
\psi(x_0)\right]e^{-ikx_0} =
\\ \nonumber
&- i\omega \frac{\Lambda}{v}u(0) - e\varphi(0) + \psi(0 ), \quad L =
\frac{i\omega + \nu}{v} \equiv \frac{i\widetilde{\omega} }{v}.
\end{align}

To ensure the validity of \Eq{eq7} in the entire complex plane of $k$, the
functions $\psi_{k}$, $u_{k}$, and $\varphi_{k}$ should have components
proportional to the factor $\exp(-ikx_0)$, which compensate the last term in
the left-hand side of \Eq{eq7}. A similar conclusion relates to the Fourier
transforms of \Eqs{eq2} and \eqref{eq3}. Thus, the system splits into two
blocks, with and without the exponential factor; however, direct application
of the Wiener-Hopf procedure to these blocks is impossible which can be
demonstrated by a simple example. The above defined Fourier transform of any
wave mode propagating with attenuation in the forward ($+$) or backward ($-$)
direction in the interval (0, $x_0$), having initial amplitudes $A_0^+ =
A^+(0)$ or $A_0^-=A^-(x_0)$, respectively, is given by
\begin{equation} \label{Apm}
A_{k}^{ \pm}  = \frac{A_0^\pm}{i(\pm r - k)}\left[e^{i(\pm r - k)x_0} - 1
\right].
\end{equation}
Here, $r$ is the complex wave number located, e.g., for the direct ($+$) wave
in the second quadrant. The Fourier components of the fields, representing a
separate solution for each block, generally contain both direct and backward
waves, therefore they have singular points in the upper and lower half-planes,
as follows from \Eq{Apm}. The Wie\-ner-Hopf method is not applicable to such
functions.

In order to get around this difficulty, we group the terms in each block
obtained from \Eq{eq7} according to the location of their singular points, i.e,
divide the full solution for each block into the forward and backward waves.
The values of the fields at the interfaces $x=0,x_0$ also contain partial
contributions of the direct and backward waves, $A(0)= A^+(0)+A^-(0)$ (and
similar for $x=x_0$), which are to be attached to the related groups. Taking
into account the existence of the common band of analyticity $-\im r < \im k <
\im r$ for these two groups and the relation ${kA_k}|_ {k \to \infty} = iA(0)$
for each partial component, and using the Liouville's theorem,\cite{Noble} we
conclude that each of these groups is equal to $0$. After such a separation,
the Wiener-Hopf method is already applicable, and we come to the conclusion
that in the case of diffusive sample boundaries, \textit{the response to an
external perturbation at the receiving interface is a combination of solutions
for the forward and backward waves in the half-space with corresponding partial
amplitudes of the perturbing signals.} The relation between these amplitudes is
to be found from the continuity conditions for the displacements and stresses
at the boundary.

By using these considerations, we address the equations for the forward wave,
obtained from the Fourier transforms of \Eqs{eq1}--\eqref{eq3} after the
separation procedure. After some al\-gebra, we get the relation between
$u_{k}$ and $\varphi_{k}$ in the forward wave (we omit the upper index $+$),
\begin{align} \label{eq8}
(k^2 - q^2)u_{k} + i\zeta ek\varphi_k \lambda_0^{-1} = - iku_0 + C_1.
\end{align}
Here and below, the symbol $C_i$ ($i = 1,2$) denotes combinations of the fields
at the exciting interface (the single used property of $C_i$ is their
independence of $k$), $\lambda _{0} = mv_F^2$, $\zeta = \lambda _{0} /Ms^2 \sim
1$, $M$ is the ion mass. Eliminating $\varphi_{k}$ from \Eqs{eq3} and
\eqref{eq8}, we arrive at the equation for $u_k$,
\begin{align}\label{eq9}
&Z(k)[kBu_k - u_0 (B+q^2) + kC_2] = A(- q^2 u_0 + kC_2)
\\ \nonumber
&- \frac{B\zeta}{\lambda_0} \left\{ \left\langle \frac{\psi_0}{L + ik}
\right\rangle_{v < 0} - \left\langle \frac{C_0}{L + ik} \right\rangle_{v < 0}
\right\}
\\ \nonumber
&Z(k) = A + BJ, \quad A = \frac{k_\omega\zeta}{3k_0}, \quad k_\omega =
\frac{\omega}{v_F}, \quad k_0 = \frac{\widetilde{\omega}}{v_F},
\\ \nonumber
&B = k^2 - q^2+k_\omega  \zeta \left(k_0 + \frac{k^2}{3k_0}\right), \quad q =
\frac{\omega}{s},
\\ \nonumber
&J = \frac{1}{\langle 1\rangle}\left\langle \frac{1}{L^2 + k^2} \right\rangle
= \frac{1}{k^2} - \frac{k_0}{2k^3}\ln\frac{k_0 + k}{k_0 - k}.
\end{align}

In derivation of \Eq{eq9}, we used the following chain of transformations,
\begin{align} \label{eq10}
&\left\langle \frac{\psi(0)}{L + ik} \right\rangle = \left\langle \frac{\psi
_0}{L + ik} \right\rangle_{v < 0} + \left\langle \frac{C_0}{L + ik}
\right\rangle_{v > 0}
\\ \nonumber
&= \left\langle \frac{\psi_0}{L + ik} \right\rangle_{v < 0} - \left\langle
\frac{C_0}{L + ik} \right\rangle_{v < 0} + C_{0} J.
\end{align}
We emphasize that the possibility to take the factor $C_{0}$ out of the
averaging in \Eq{eq10} determines the applicability of the Wiener-Hopf method
to our problem. We also note that the combination in the curly brackets in
\Eq{eq9} plays the role of a ``fictitious'' function appearing in this
method.

The characteristic function $Z(k)$ determines the spectrum of the wave
numbers of the propagating modes. In our simplest case, the equation $Z(k) =
0$ has only a pair of the roots $r_\pm  = \mp q \pm i\alpha_L$,
corresponding to the acoustic wave renormalized by interaction with
electrons. At $q\ell \gg 1$, the attenuation decrement $\alpha_L = \pi
k_\omega/12$ represents the Landau damping independent of the mean free path
$\ell = v_{F}/\nu$. Besides, the function $Z(k)$ has a pair of the branch
points $k = \pm k_0$ associated with the quasiwave (ballistic) process with
the propagation velocity close to $v_F$. The function $Z(k)$ has no singular
points near the real axis within the band $ -\delta < \im k < \delta$,
$\delta = \text{min}(\alpha_L, \ell^{-1})$, and turns to unity at $k \to
\infty$. These properties enable us to factorize $Z(k)$ by using a standard
procedure,\cite{Noble} i.e., to present it as a product of the functions
$T^{+}(k)$ and $T^{-}(k)$, analytical at $\im k
> -\delta$ and $\im k < \delta$, respectively. In particular,
\begin{align} \label{eq11}
T^+(k) = \exp\left[\frac{1}{2\pi i}\int_{ - \infty - i\gamma}^{\infty -
i\gamma} \frac{\ln Z(\xi)}{\xi - k} d\xi \right],\quad \gamma < \delta.
\end{align}

This function can be calculated by the methods of contour integration. In the
lower half-plane, the integrand in \Eq{eq11} has two branch points: $\xi =
k_0$ from the internal logarithm in $Z(\xi)$ and $\xi = r_-$, in which the
function $Z(\xi)$ turns to zero. We make a cut for the internal logarithm
along the ray $\xi = k_0y$ ($1 < y < \infty$). Since the function $Z(\xi)$ is
regular at $\xi = 0, \infty$, the second cut, beginning at the point
$\xi=r_-$, is finished at some point $\xi=r_0$ belonging to the first cut.
Then, tracing the cuts and calculating corresponding contour integrals, we
get
\begin{equation} \label{eq12}
T^+(k) = \frac{k - r_-}{k - r_0}\tau^+(k),
\end{equation}
where $\tau^+(k)$ is the contribution of the cut of the internal logarithm,
\begin{equation}\label{eq13}
\tau^+ (k) = \exp\Bigl[ \frac{k_0}{2\pi i}\int_{1}^{\infty} \frac{\ln Z(k_0 y +
0i)-\ln Z(k_{0} y - 0i)}{k-k_0 y} dy \Bigr].
\end{equation}
The value of $r_0$ can be found assuming $k = 0$ in \Eq{eq11}. In this
limit, after displacement of the integration contour to the real axis, the
principal value of the integral vanishes and only the contribution $\pi i
\ln Z(0)$ of the trace around the coordinates origin survives. Comparing
this result with \Eq{eq12}, we get $r_{0} = r_- \tau^+(0)Z^{- 1/2}(0)$. Note
that the formal singularity in \Eq{eq12} at $k = r_0$ is removable, because
$\tau ^{+}(k)_{k \to r_0} \to 0$.

Dividing \Eq{eq9} over $T^+(k)$, we obtain a functional equation with the
right-hand and left-hand sides analytical at $\im k > -\gamma$ and $\im k <
\gamma$, respectively. Due to the Liouville's theorem, they can be presented as
a first power polynomial $ A(\alpha k + \beta)$, and we obtain the final
expression for the Fourier image of the elastic displacement in the forward
wave:
\begin{equation} \label{eq15}
u_k = \frac{A(\alpha k + \beta)T^+ (k)}{ikB(k)Z(k)}.
\end{equation}
According to \Eq{eq5}, the combination in curly brackets in \Eq{eq9} vanishes
at $k = 0$, which gives $\beta = \sqrt {3} qk_0 u_0$. Expression for the
parameter $\alpha$ follows from \Eq{eq9} at $k = \pm b$, where $b$ is a root
of the equation $B(k)=0$. Note that we omitted the terms in \Eq{eq15}, which
are inessential for the calculation of $u(x)$ [they eliminate nonphysical
poles in \Eq{eq15} at $k = 0,\pm b$ and give zero contribution to the inverse
Fourier transform of $u_k$].

According to \Eq{eq15} and the properties of the function $Z(k)$ described
above, the forward wave consists of two modes having different velocities: the
quasiwave and the sound wave. Elastic displacements in these excitations,
$u_{\text{qw}} \sim (s/v_F)^2 u_0$ and $u_q \sim u_0$, respectively, differ by
4--5 orders of magnitude. However, the electric potentials excited by these
modes are comparable by their magnitudes,
\begin{equation} \label{Phi}
\varphi_{\text{q}} \sim \varphi_{\text{qw}} \sim \frac{k_\omega
\lambda_0}{e}u_0.
\end{equation}
These estimations follow from \Eq{eq8} at $k=q+i\alpha_L$, $u \sim u_0$ for
the sound wave and $k \approx k_\omega$, $u \sim (s/v_F)^2 u_0$ for the
quasiwave. Comparing \Eq{Phi} with \Eq{eq4}, we conclude that the resulting
sound potential is formed by practically complete ($\sim s/v_F$)
cancellation of two large terms. On the contrary, the quasiwave potential is
mainly represented by the electric component.

Now we consider the values of the fields registered at the receiving interface.
The excitation incoming at the sample boundary $x=x_0$ produces its
deformation, which is the source of backward waves. Their behavior is described
by similar equations with substituting $u_0$ by the partial amplitude
$\widetilde{u}(x_0)$, whose value can be found from the mechanical boundary
conditions. The measured potential is the sum of contributions of the forward
and the backward waves. As an example, we find the elastic displacement and the
potential created by a direct quasiwave, coming with the amplitude
$u_{\text{qw}}(x_0)$ to the matched interface. In this case, evaluation of the
integral in \Eq{eq13} is required; however, we will use instead the
characteristic values of $k \sim k_0 \ll q$ for estimations. For the backward
waves, the quasiwave contribution is negligible, and we should take into
account only the acoustic component. In this approximation, the conditions of
equality of displacements and stresses for both sides of the interface read as
\begin{align} \label{eq16}
&u_{\text{ES}} = u_{\text{qw}} + \widetilde{u}(x_0), \quad \widetilde{u}(x_0)
\approx \frac{1}{2iq}\frac{W_{\text{qw}}(x_0)}{\rho s^2} -
\frac{u_{\text{qw}}}{2},
\\
& -iqu_{\text{ES}} = ik_0 u_{\text{qw}}(x_0) + iq\widetilde{u}(x_0) -
\frac{W_{\text{qw}}}{\rho s^2}.
\end{align}
Here $u_{\text{ES}}$ is the amplitude of displacements created by the electron
sound signal in a load (including the receiving piezotransducer). The
electronic pressure for a direct quasiwave can be found from the Fourier
transform of \Eq{eq2},
\begin{equation}\label{eq17}
\frac{W_{\text{qw}}}{\rho s^2} \approx \frac{q^2 - k^2}{ik}u_{\text{qw}}(x_0).
\end{equation}
As a result, we obtain $u_{\text{ES}} \approx \widetilde {u}(x_0) \approx
({q}/{2k_0}) u_{\text{qw}}(x_0) \sim ({s}/{v_F})u_0$, i.e., the displacement
amplitude at the receiving interface exceeds its value in the incoming wave of
the electron sound by a large factor $v_F/s$.\cite{Avr} At the same time, the
potential created by the backward waves is small, thus, $\varphi_{\text{ES}}$
equals to the potential of a direct quasiwave $\varphi_{\text{qw}}(x_0)$.
Although the quantity $\widetilde{u}(x_0)$ at the free boundary is twice as
large, the contribution of backward waves can be also neglected in this case.
This means that the amplitude of $\varphi_{\text{ES}}$ is practically
independent of the mechanical boundary conditions, in agreement with our
experiments.

For the sound potential, the cases of the matched and the free boundaries
differ in essence. In the first case, the quantity $\widetilde{u}(x_0)$ is
small by the parameter $\alpha_{L}/q$, thus the contribution of the secondary
waves is negligible, and $\varphi_{\text{S}}(x_0)$ equals to the potential of
the primary sound wave. For the free interface, $\widetilde{u}(x_0)$ coincides
with the amplitude of the incident wave, their potentials are fully
compensated, and only the potential of the secondary quasiwave survives. In the
regime of the impurity scattering (low-temperature limit), the quasiwave
contribution exceeds the acoustic one by the factor of 1.5 (for the specular
boundary,\cite{AvrGokh} $\varphi_{\text{qw}}$ exceeds $\varphi_{q}$ by a factor
more than 3). When the electron scattering increases, the acoustic component
$\varphi_{q}$ always becomes prevalent.

The experimental dependencies in Fig.~\ref{fig2} qualitatively agree with the
estimations given above. Of course, the used variants of measurement of the
potential cannot be attributed to the purely matched or free boundary,
therefore both the acoustic and quasiwave contributions are present in
$\varphi_{\text{S}}$. Nevertheless, we note that in the regime of the impurity
scattering, $\varphi_{\text{S}}$ is larger for the variant more close to a free
boundary case than for a matched one. And vice versa, in a high-temperature
region, only $\varphi_{\text{q}}$ remains, and a more intensive signal is
observed in the ``matched'' variant.

Within the same model, we analyze the case of a specular receiving interface,
assuming the exciting interface to be diffusive to avoid possible resonant
effects. Different authors used various approaches to similar problems (see,
e.g., Ref.~\onlinecite{Abrikosov}) but they actually exploit an identical
procedure --- replacement of the interface by a specularly inverted sample. The
scalars in the fictitious sample are the same as in the real one, the
$x$-components of polar vectors change their signs, and the tensor functions
are transformed in accordance with the usual rules. Due to the specular
reflection conditions, the distribution function is continuous at $x=x_0$.

Applying the Fourier transformation [now within the interval (0, $2x_0$), out
of which all fields are assumed to be equal to $0$] to our system, we conclude
that the complete solution splits into three blocks [cf.~with \Eq{eq7}]. One of
them (without the exponential factor) coincides with the one discussed above
and describes the waves generated at the interface $x = 0$. Two others [$\sim
\exp(- 2ikx_0)$ and $\sim \exp(- ikx_0)$] are virtual excitations, but their
sum determines real backward waves. The first of these terms is the
``specularly inverted'' wave, in accordance with the rules accepted. Obviously,
at $x = x_0$, this wave produces a displacement opposite in phase and a
potential of the same sign compared to those in the incoming wave. The second
term is the excitation generated by complete (not partial!) displacements
$u(x_0)$ of the interface. In our notations, its Fourier transform is
\begin{equation} \label{eq18}
u_k = - \frac{2Aq^{2}u(x_0)}{ikB(k)Z(k)},
\end{equation}
where we omitted inessential terms, which cancel out the poles at $k = 0,
\pm b$, similar to \Eq{eq15}. The potential is determined by \Eq{eq8} with
minor modification of the right-hand side. If one considers $u(x_0)$ as an
independent value, then \Eqs{eq15} and \eqref{eq18} determine the
relationship between the amplitudes of displacements generated at the
diffusive and specular boundaries, respectively. For the acoustic mode, it
is very close to 1, while for the quasiwave this relation is about $0.5$.

In analysis of the mechanical boundary conditions, all three solutions must be
taken into account. Obviously, the sum of first two solutions gives zero
displacement and doubled potential and electronic pressure. In the case of a
quasiwave incident on the specular interface, the corrections arising from the
backward waves are small, therefore the full potential $\varphi_{\text{ES}}$ at
the specular interface is twice as large than at the diffusive one.

In the case of the sound wave incident on the matched interface, the quantity
$u(x_0)$ coincides with the incoming signal. Summing up the solutions, we
find the potential created by the acoustic wave and an additional potential
generated by the quasiwave. The relationship between these contributions
coincides with that calculated before.\cite{AvrGokh} The amplitude of
displacements in the backward sound wave is small because of practically full
cancellation of the second and third terms.

Thus, for the specular interface, the potentials $\varphi_{\text{ES}}$ and
$\varphi_{\text{S}}$ exceed, as a rule, their values for the diffusive case.
The only exception is a hypothetical fully fastened surface, $u(x_0) = 0$; in
the diffusive case, both the doubled potential $\varphi _{q}$ and $\varphi
_{\text{qw}}$ contribute to $\varphi_{\text{S}}$, while for the specular
interface, the term $\varphi _{\text{qw}}$ is absent. It is also worth noticing
that there is a qualitative difference between the diffusive and specular cases
for the matched interface in the clean limit $q\ell \gg 1$: the contribution to
$\varphi_{\text{S}}$ from the quasiwave for a diffusive boundary is practically
absent, while it dominates in a specular case.\vspace{-0.5cm}

\subsection{Multiband models}\vspace{-0.3cm}

Despite the successful explanation of several important experimental facts, the
free-electron model has an essential drawback: it does not explain the
difference between the phases of $\varphi_{\text{ES}}$ and $u_{\text{ES}}$
clearly seen in Fig.~\ref{fig1}(b). Indeed, comparing \Eq{eq8} and \Eqs{eq16},
\eqref{eq17}, we see that the electron sound potential and elastic
displacements at the interface are described by expressions, similar up to a
scale factor. It seems that the consideration of the quasiwave as a single
carrier of the electron sound will lead to an analogous conclusion for any
modification of the approach.

However, the quasiwave is not a unique mechanism of the electron sound
transport. In the presence of strong enough Fermi-liquid interaction (FLI) and
several sheets of the Fermi surface with close Fermi velocities but different
values of the deformation potential, the excitation of zero sound in metal is
possible.\cite{Dunin,Dubovik,ZS,BezPhysicaB,BezCondMatter,BezJLTP} It was
found\cite{BezVelocity,Avr} that a considerable change in the phase of the
elastic component of the electron sound in Ga with temperature is related to
the change of its velocity, associated with the crossover from the
collisionless propagation of the zero sound to the concentration wave
regime\cite{BezVelocity,Kopel} (the electron analog of ordinary sound).
Theoretical analysis, based on the model of a compensated metal with two
equivalent zones, showed that the necessary condition for such a crossover is
relatively weak interband scattering.\cite{BezVelocity} This requirement is not
an artificial limitation of the model, since the interband gaps are often large
enough, therefore in the actual range of temperatures, the interband
transitions are only due to rare electron-impurity or electron-electron
collisions. At the same time, the intraband relaxation above the crossover
temperature is determined by much more frequent electron-phonon collisions.

Within this model, the elastic component of the zero sound (or concentration
mode) predominates the ballistic one at reasonable values of the FLI
parameters, but the potentials $\varphi_{\text{ES}}$ and $\varphi_{\text{S}}$
are identically zero. For their emergence, a certain asymmetry must be
introduced: unequal FLI coefficients, different densities of states, or
different (but close) values of the Fermi velocities. However, in this case,
the phase of the zero sound potential behaves similar to the phase of the
elastic component. Thus, the two-band model also cannot give any explanation of
the data presented in Fig.~\ref{fig1}(b).

A qualitative interpretation of these data can be obtained within the framework
of a three-band model. We represent the Fermi surface by three spheres of
identical sizes, two of which are of the electron type and one of the hole type
(or vice versa). The Fermi velocities, the densities of states, the relaxation
rates, and the intensity of FLI are supposed to be equal for all bands.
Besides, we assume the absence of interband transitions caused by the
electron-phonon scattering and equality of the rates of the intra- and
interband impurity scattering. Under these assumptions, the kinetic equation in
each band ($i = 1,2,3$) for the distribution function renormalized by
FLI\cite{Avr,BezVelocity} has a form similar to \Eq{eq1}, with an additional
force term in the right-hand side,
\begin{equation} \label{eq19}
i\widetilde{\omega} \psi_i + v\frac{d\psi_i }{dx} + \nu \psi = - i\omega
\Lambda_i \frac{du}{dx} + ev_i \frac{d\varphi}{dx} + \frac{\omega^-
}{\left\langle {1} \right\rangle}\left\langle{\psi_i} \right\rangle,
\end{equation}
where $F$ is the difference of the isotropic parts of Landau correlation
functions for the intra- and interband FLI, $\omega ^{ -}  = \nu _{ph} +
i\omega F/(1 + F)$, $\nu = \nu_{ph} + 3\nu_{imp} $, $\nu_{ph} $ and $\nu
_{imp}$ are the frequencies of the intraband electron-phonon and
electron-impurity collisions, respectively. The FLI renormalizes the function
$W$ in \Eq{eq2} as well,
\begin{equation}\nonumber
W = \sum\nolimits_{i}\left[ {\left\langle {\Lambda _{i} \psi _{i}}
\right\rangle} - \frac{{F}}{{1 + F}}\left\langle {\Lambda _{i}} \right\rangle
\frac{{\left\langle {\psi _{i}}  \right\rangle} }{{\left\langle {1}
\right\rangle} }\right],
\end{equation}
and \Eqs{eq3} and \eqref{eq5} take the form $\sum\nolimits_i \langle \psi_i
\rangle =0, \quad \sum\nolimits_i \langle v_i\psi_i \rangle =0$.

\begin{figure}[tb]
\centerline{\epsfxsize=8.5cm\epsffile{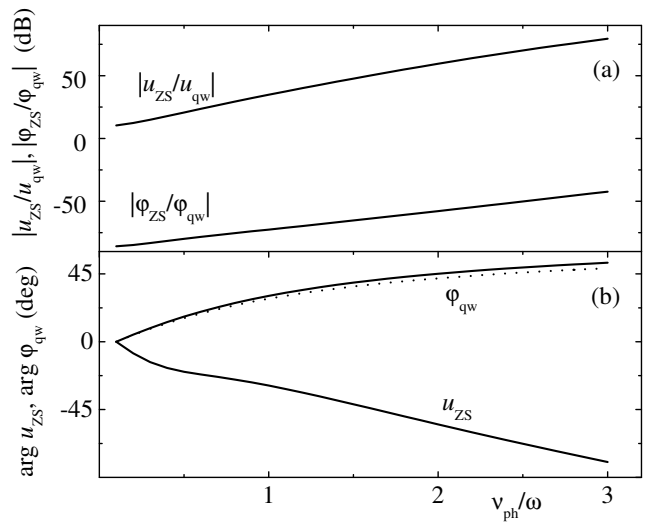}}
\caption{Calculated relations (a) of the amplitudes of elastic displacements
and potentials for separate components of electron sound vs electron-phonon
scattering rate in the three-band model and (b) of computed phases of the
dominant components. The following parameters are used: $\zeta = 1$, $F = 1$,
$\omega/3\nu_{imp} = 5$. Dotted line is the phase of the quasiwave components
for the free electron model.}\vspace{-0.5cm}
\label{fig3}
\end{figure}

As well as in the two-band model,\cite{Dunin,Dubovik,BezPhysicaB} these
equations have a zero-sound solution transformed into the concentration mode
with the increase of scattering. The results, obtained for the specular
receiving interface, show that for reasonable intensity of FLI ($F \sim 1$),
the elastic and potential components of the electron sound in this case are
formed by different mechanisms. Indeed, as is obvious from Fig.~\ref{fig3}(a),
the elastic component $u_{\text{ZS}}$ of the zero sound much exceeds its value
in the quasiwave and, at the same time, the quasiwave potential dominates. As a
result, the behavior of the phases of dominant components presented in
Fig.~\ref{fig3}(b) qualitatively agrees with the experimental data shown in
Fig.~\ref{fig1}(b).

The behavior of the phase of the quasiwave potential, following from the
one-band model with the diffusive surface, is also presented in
Fig.~\ref{fig3}(b). Almost complete coincidence of these results with the ones
for the three-band model indicates insensitivity of the phase of quasiwave
solutions to the particular choice of the model and to the character of
electron scattering at the interface.\vspace{-0.5cm}

\section{Behavior of potential in superconducting phase}\vspace{-0.3cm}

The algorithm of calculation of the potential in the super\-conducting phase is
similar to the procedure described in Sec.~III. In particular, the same
equations of elasticity, electro- and current neutrality are used. Of course,
the calculation of corresponding averages is much more difficult due to the
energy dependence of both the velocity of normal excitations and the relaxation
frequencies.\cite{AvrGokh,Boichuk} However, in derivation of \Eq{eq8} within
the free-electron model, no specific calculations of the kinetic coefficients
were performed, therefore its structure holds in the superconducting state as
well. This means that $\varphi_\text{q}$ cannot decrease below $T_\text{c}$
faster than the sound attenuation decrement $\alpha_L(T)$ (for the case of a
specular interface, a detailed analysis was given in
Ref.~\onlinecite{AvrGokh}). Moreover, since the sound attenuation in our sample
is rather large, $\alpha_L x_0 > 1$, the dependence $\varphi_{\text{q}}(T)$
must pass through a maximum due to rapid increase of the damping factor
$\exp[-\alpha_L(T) x_0]$ near $T_{\text{c}}$. The relationship similar to
\Eq{eq8}, following from the elasticity equation, occurs in any model,
therefore the conclusion about the temperature dependence of $\varphi_\text{q}$
seems to be always true.

\begin{figure}[tb]
\centerline{\epsfxsize=8.5cm\epsffile{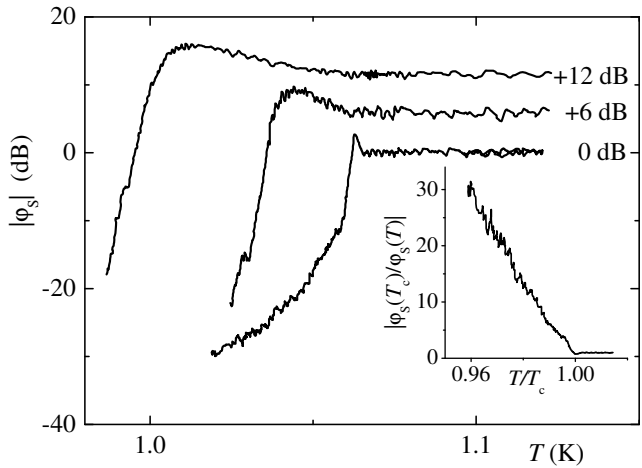}}\vspace{-0.3cm}
\caption{Amplitude of the sound wave potential $\varphi_{\text{S}}$ vs
temperature below $T_{\text{c}}$ at different levels of the exciting signal.
Inset: behavior of $|\varphi_{\text{S}}(T)|^{-1}$ in a near vicinity of
$T_{\text{c}}$ for the curve 0 dB.}\vspace{-0.3cm}
\label{fig4}
\end{figure}

However, as it has been already reported,\cite{AvrGokh} the experimental value
of $\varphi_{\text{S}}$ decreases considerably faster than expected on the
basis of these considerations. We note that the measurement of the potential in
these experiments was carried out by a point contact, for which the mechanical
boundary conditions depend on its pressing, i.e., on a badly controlled
parameter. In particular, it can be thought that the situation with a point
contact is close to the case of free boundary, and, correspondingly, the
contribution of $\varphi_{\text{q}}$ in the potential measured in
Ref.~\onlinecite{AvrGokh} is completely absent. In the present experiments,
$\varphi_{\text{q}}$ is unambiguously a part of $\varphi_{\text{S}}$;
nevertheless, the result of measurements of $\varphi_{\text{S}}(T)$ in the
superconducting state shown in Fig.~\ref{fig4} completely reproduces the
previous result.

At large excitation intensity, $|\varphi_{\text{S}}(T)|$ exhibits a maximum,
which is due to local overheating of the receiving interface and vanishes with
the decrease of $u_{0}$. In the absence of the overheating, the quantity
$|\varphi_{\text{S}}(T)|^{ - 1}$ obeys the law close to linear in $\Delta T =
T_{\text{c}}- T$ (Fig.~\ref{fig4}, inset) with a large prefactor similar to
that in the imaginary part of the transversal conductivity of a superconductor,
$\im\sigma_s/\sigma_n \approx (2v_F/s)(\Delta T/T_{\text{c}})$, which describes
screening of the electromagnetic field of the sound wave by supercurrents. This
enables one to suspect that the oscillating currents, spreading over the sample
surface from the sound spot, take part in formation of $\varphi_{\text{S}}(T)$.
These currents were indeed observed in the experiments;\cite{AvrGokh} however,
they disappear at $T<T_{\text{c}}$ as quickly as the potential does and
therefore hardly can be a primary cause of its decrease.

\begin{figure}[tb]
\centerline{\epsfxsize=8.5cm\epsffile{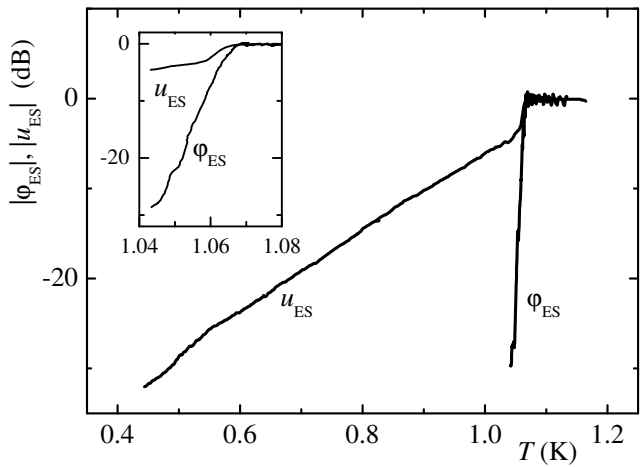}}\vspace{-0.3cm}
\caption{Changes of the amplitudes of the potential and elastic displacement in
the electron sound wave below $T_{\text{c}}$. Inset: expanded scale near
$T_{\text{c}}$.}\vspace{-0.5cm}
\label{fig5}
\end{figure}

Generally, the theoretical analysis of $\varphi_{\text{S}}(T)$ and the
interpretation of its experimental behavior represent a rather complicated
problem, because both $\varphi_\text{q}$ and $\varphi_{\text{qw}}$ contribute
to the sound potential. In principle, these terms may compensate each other,
although such a situation seems to be hardly probable. Moreover, as was noted
above, the potential $\varphi_\text{q}$ is the result of practically complete
($\sim s/v_F$) cancellation of large electric and deformation contributions;
for this reason, the ordinary accuracy of estimations (also $\sim s/v_F$) must
be substantially increased. In this sense, the measurement of
$\varphi_{\text{ES}}$ is more preferable, because this potential has a purely
electric nature. The results of measurements of $\varphi_{\text{ES}}(T)$ and
$u_{\text{ES}}(T)$ presented in Fig.~\ref{fig5} show that the potential of the
electron sound disappears at $T<T_{\text{c}}$ practically in a jumplike way,
similar to $\varphi_{\text{S}}$. Strangely, but the result of measuring
$\varphi_{\text{S}}$ and $\varphi_{\text{ES}}$ looks as an evidence of
impossibility of the existence of the potential gradient in a superconductor.
Of course, we do not adhere to such a point of view, because it fully
contradicts the universally recognized theories and well-established
experimental facts (see, e.g., a review\cite{Artemenko}), but the problem of
interpretation of these paradoxical data still exists.

The nature of the signal $u_{\text{ES}}(T)$ also remains unclear. Taking into
account the analysis of the three-band model, it could be thought that a small
jump in $u_{\text{ES}}(T)$ near $T_{\text{c}}$ (see Fig.~5) can be interpreted
as the suppression of the quasiwave just below $T_{\text{c}}$. Furthermore, it
was shown experimentally\cite{Avr} that the change of both the amplitude and
the phase of $u_{\text{ES}}(T)$ below $T_{\text{c}}$ has nothing to do with the
change of attenuation and velocity of the electron sound and relates only to
the behavior of the coefficient of coupling between the electron sound and the
exciting elastic deformation. This contradicts the theoretical
predictions\cite{Boichuk} about the behavior of the quasiwave amplitude and
phase in the superconductor. Besides, we would remind the conclusion of
Ref.~\onlinecite{Leggett} that in presence of the interband Cooper pairing, the
zero sound spectrum in the superconducting phase has an activating character
with a gap close to the energy gap of the superconductor. Thus, the propagation
of the zero sound at our frequencies is forbidden in the superconducting state.
But if the signal $u_{\text{ES}}(T)$ below $T_{\text{c}}$ is neither zero sound
nor the quasiwave, then what is it? No clear answer on this question exists
yet.\vspace{-0.5cm}

\section{Conclusion}\vspace{-0.3cm}

We have measured the temperature dependencies of the amplitude and the phase of
the potential $\varphi_{\text{ES}}$ and the elastic dis\-placement
$u_{\text{ES}}$ accompanying a fast electron sound wave excited by the
longitudinal ultrasound in a single crystal of high-purity Ga. Simultaneously,
the amplitude and the phase of the potential $\varphi_{\text{S}}$ and the
elastic displacement $u_{\text{S}}$ in the excited acoustic wave have been
studied. We found that in the normal state, the behavior of the phases of
$\varphi_{\text{ES}}$ and $u_{\text{ES}}$ differs qualitatively: while the
phase of $\varphi_{\text{ES}}$ increases with temperature, the phase of
$u_{\text{ES}}$ decreases. By using the Wiener-Hopf method, we examined several
theoretical models that describe excitation and propagation of different types
of the electron sound in samples of finite size with the diffusive exciting
interface.

The model of free electrons, in which only the quasiwave is responsible for the
electron sound transport, enabled us to explain several important experimental
facts: giant enhancement (by the factor $v_F/s$) of elastic displacements
induced by the electron sound wave at the sample boundary, insensitivity of
$\varphi_{\text{ES}}$ on the boundary conditions at the receiving interface,
the temperature behavior of $\varphi_{\text{S}}$ and its closeness to
$\varphi_{\text{ES}}$ in the magnitude. However, neither this model nor the
model of a compensated metal with two sheets of the Fermi surface (in which
zero-sound or concentration modes occur in presence of the Fermi-liquid
interaction) are able to explain the difference between the phases of
$\varphi_{\text{ES}}$ and $u_{\text{ES}}$.

We obtained a qualitative interpretation of this experimental result within a
model with three equal Fermi spheres, which reflects the presence of three main
sheets of the Fermi surface in Ga.\cite{Reed} For reasonable values of the
Fermi-liquid interaction coefficients, the elastic signal $u_{\text{ES}}$ was
found to be formed by the zero sound, while the potential $\varphi_{\text{ES}}$
is basically associated with the quasiwave, which results in opposite changes
of their phases with temperature. Of course, this simple model cannot pretend
to be a quantitative description of the real situation. Nevertheless, our
estimations indicate a possibility, in principle, for the ``potentialless''
propagation of the zero sound (or the concentration wave) on a background of
the potential created by the ballistic transport and enable us to suppose
actual realization of a similar scenario in the experiments.

Below the temperature of the superconducting transition, we observed a sharp
disappearance of the potential of both the electron sound wave and the acoustic
wave, which contradicts our theoretical estimations and generally adopted
conceptions of the behavior of the longitudinal electric field in
superconductors. The origin of this puzzling effect, as well as the nature of
the elastic signal of the electron sound in the superconductor, is not clear
yet.

The authors are thankful to L.~A.~Pastur and D.~V.~Fil for stimulating
discussions.

\end{document}